\documentclass[%
aip,
amsmath,amssymb,
preprint,%
]{revtex4-2}

\usepackage{graphicx}
\usepackage{dcolumn}
\usepackage{bm}

\usepackage{bbding}
\usepackage[utf8]{inputenc}
\usepackage[T1]{fontenc}
\usepackage{mathptmx}
\usepackage{etoolbox}

\makeatletter
\def\@email#1#2{%
	\endgroup
	\patchcmd{\titleblock@produce}
	{\frontmatter@RRAPformat}
	{\frontmatter@RRAPformat{\produce@RRAP{*#1\href{mailto:#2}{#2}}}\frontmatter@RRAPformat}
	{}{}
}%
\makeatother

\newcommand{\bea}{\begin{eqnarray}}
\newcommand{\eea}{\end{eqnarray}}
\newcommand{\be}{\begin{equation}}
\newcommand{\ee}{\end{equation}}
\newcommand{\eps}{\varepsilon}

\begin{document}
	
	
	\title[Negative/positive electrocaloric effect in antiferroelectric squaric acid]{Negative/positive electrocaloric effect in antiferroelectric squaric acid}
	\author{A.P. Moina}
	\affiliation{ 
Institute for Condensed Matter Physics,  1	Svientsitskii St., 79011 Lviv, Ukraine\\
	alla@icmp.lviv.ua}%

	\date{\today}
	
	\begin{abstract}
		Using the previously developed model we study the electrocaloric effect (ECE) in antiferroelectric crystals of squaric acid. At low temperatures, the polarization reorientation by the electric field applied along the crystallographic $a$ axis of these crystals is predicted to be a two-stage process, with one of the sublattice polarizations switched twice, by 90$^\circ$ each time. The $T$-$E$  landscapes of the model polarization and entropy are explored. The ECE, characterized by the introduced electric Gr\"uneisen parameter, is found to be negative  in the antiferroelectric phase. In the intermediate ferrielectric phase with perpendicular sublattice polarizations it is positive at low fields, but becomes  slightly negative below the transition to the ferroelectric phase. In the ferroelectric phase the ECE is positive at all temperatures and fields. Maximal EC temperature shift magnitude is predicted to be just less than $-3$~K at 200~kV/cm. The supercritical behavior of the Gr\"uneisen parameter in the crossover region between two
		bicritical end points is explored.
	\end{abstract}
	
	\maketitle

\section{Introduction}
The electrocaloric effect (ECE) is a change of temperature or entropy of a sample by varying external electric field. The ECE observed in ferroics is believed to be promising for potential applications in solid-state refrigeration, as the experimentally obtained temperature shifts in these materials can be quite significant (see the reviews \cite{scott:16,moya:14,kutnjak:15}).

Normally, the external field applied to a dielectric orders the electric dipoles in the system, thereby decreasing its entropy, or, if the process is adiabatic, increasing its temperature. This is the conventional or positive ECE ($dT/dE>0$). However, if the initial ordering of the dipoles is antiferroelectric (AFE), then the electric field tends to destroy it and increases the system entropy. In this case the ECE is negative or inverse, and the sample temperature is decreased ($dT/dE<0$).
This and other circumstances, when the ECE in ferroics is negative, are substantially discussed in the review \cite{grunebohm:18}.
Commonly, the negative ECE in antiferroelectrics coexists with the positive one, observed in other field or temperature domains.  Those are, for instance, the temperatures above the N\'{e}el temperature and the fields above the field-induced transition from the antiferroelectric  to ferroelectric (FE) phase, at which  one of the sublattice polarizations is switched by  $180^\circ$, from the antiparallel position to the parallel one.

The magnitude of the EC temperature shift  $\Delta T$ is expected to increase in the vicinities of phase transitions, where the variation of entropy and polarization should be most prominent.
If a line of the phase transitions in the $T$-$E$ plane exhibits a multicritical point, the largest $\Delta T$ often
occurs near that point \cite{pirc:14,squillante:21,kutnjak:15}.
In antiferroelectrics,  the line of the AFE-FE transition is of a special interest in this context. As follows from the analysis of the simplest Kittel model, this line, terminating at the N\' eel temperature at zero field, does have a tricritical point at the used in calculations values of the Landau coefficients, and  $\Delta T$ is indeed the largest in its vicinity \cite{pirc:14,kutnjak:15}.

Squaric acid, H$_2$C$_4$O$_4$, is a classical two-dimensional antiferroelectric. However, some unique peculiarities of the process of the AFE-FE switching in it render the ECE in squaric acid particularly interesting to explore.

\begin{figure}[hbt]
	\includegraphics[width=0.7\textwidth]{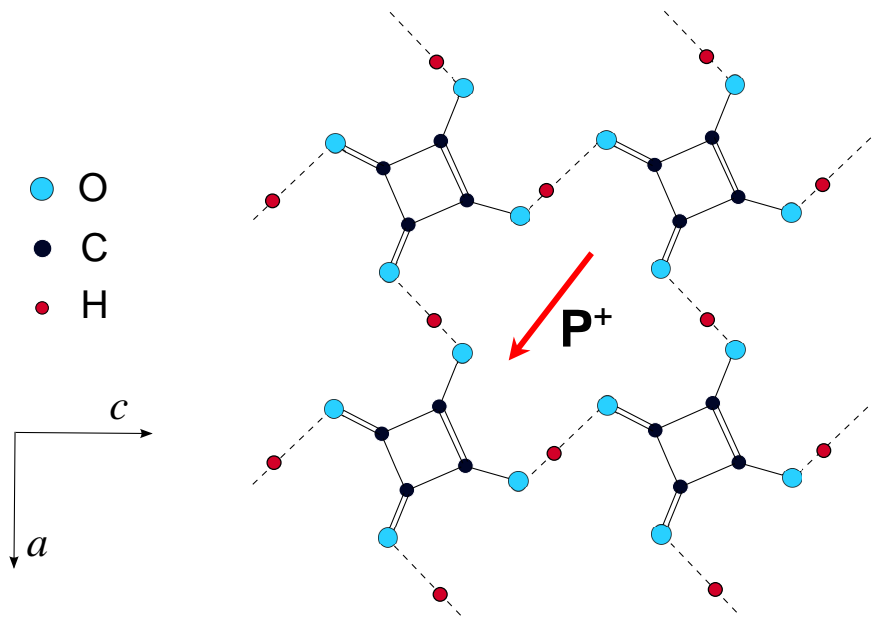}
	\caption{The pseudotetragonal fully ordered AFE structure of a sheet of H-bonded C$_4$O$_4$ groups in the squaric acid, drawn using the neutron diffraction data \cite{semmingsen:95} (proton off-center displacements are exaggerated).  Two adjacent protons near each group are localized in the vicinal to this group sites on the H-bonds, and the two others in the distal sites (a lateral configuration). Note a correlation between the positions of the double bonds and of the vicinal protons. One of four ground state configurations is shown; the three others can be obtained by rotating the entire picture by a multiple of 90$^\circ$. Direction of the sublattice polarization $\textbf{P}^+$  for the shown configuration and the crystallographic axes $a$ and $c$  are indicated.}
	\label{structure}
\end{figure}

The  high-temperature phase of squaric acid is tetragonal ($I4/m$), whereas the low-temperature structure, formally monoclinic ($P2_1/m$), is pseudotetragonal \cite{semmingsen:77}. Parallel to the $ac$ plane sheets of C$_4$O$_4$ groups, linked by hydrogen bonds along two perpendicular directions as shown in fig.~\ref{structure}, are stacked along the unique $b$ axis. Below the transition temperature, 373~K, a ferroelectric polarization arises in each sheet, and the neighboring ones polarize alternatively in the opposite directions.

\begin{figure}[hbt]
	\includegraphics[width=0.4\textwidth]{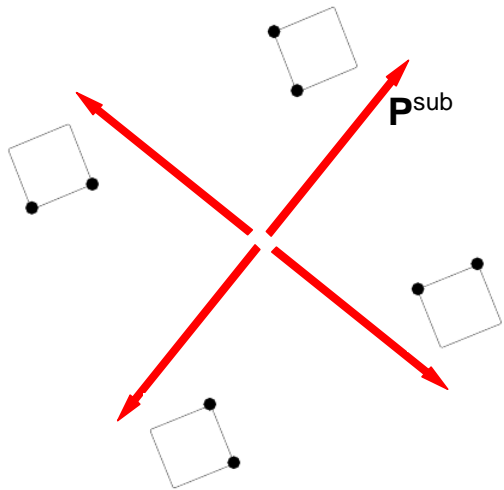}
	\caption{The allowed directions of the sublattice polarization
	in fully ordered squaric acid and the schemes of the associated lateral proton configurations. Squares are the C$_4$O$_4$ groups, and black circles are the vicinal H-sites, occupied by protons. }
	\label{twoaxis}
\end{figure}

Protons in squaric acid move in double-well potentials between
two off-center cites on the H-bonds. The sublattice polarizations are formed \cite{horiuchi:18} by the protonic displacements and, most importantly, by switchable $\pi$-bond dipoles, positions of which in a given C$_4$O$_4$ group are determined by the proton arrangement around it  (see fig.~\ref{structure}). Because of the pseudotetragonal symmetry of the hydrogen bond network, the resulting ground state sublattice polarizations can be oriented along one of two perpendicular axes, depending on which two adjacent vicinal H-sites are occupied by the ordered protons (fig.~\ref{twoaxis}). These  polarization axes, %
most likely, run close to the diagonals of the $ac$ plane \cite{horiuchi:18,horiuchi:21}.

It has been suggested \cite{horiuchi:18,ishibashi:18} that, as two   polarization axes are present, the external electric field applied within the $ac$ plane can switch the sublattice polarization  not only by  180$^\circ$ at once, but also by two consecutive (at different field magnitudes) steps of 90$^\circ$. In this case, between the initial AFE and field-induced collinear FE phases, the intermediate noncollinear phase with perpendicular sublattice polarizations (NC90) should appear \cite{moina:21}. Switching of the fully ordered sublattice polarization by 90$^\circ$ requires relocation of two protons around \textit{each} C$_4$O$_4$ group to the other potential wells within the respective H-bonds, causing also switching of two $\pi$-bonds to other C and O atoms of the group \cite{horiuchi:18,moina:21}. Switching by
180$^\circ$ requires all four protons around each group to be relocated and three $\pi$-bonds to be switched simultaneously, which is apparently more energy consuming. Due to thermal fluctuations and spontaneous switching of dipole moments of \textit{some} H$_2$C$_4$O$_4$ groups in presence of the bias field, the resulting angle between the sublattice polarizations, as statistically averaged quantities, may, in fact, be arbitrary at the elevated temperatures, and not just 0/90/180$^\circ$.

The deformable two-sublattice proton ordering model \cite{moina:20,moina:21,moina:21:2,moina:22} has been proposed in order to explore the electric field effects in squaric acid. 
The model calculations predict \cite{moina:21,moina:21:2} that 
the polarization switching at low temperatures is the two-step 
process for all orientations of the external field within the $ac$ plane, except for some particular ones: i) when the field is directed along the 
axis of spontaneous sublattice polarizations, the intermediate noncollinear phase is absent, and the switching occurs as in usual uniaxial antiferroelectrics by 180$^\circ$ at once, and ii) when the field is at 45$^\circ$ to this axis, the field of the transition to the ferroelectric phase tends to infinity. 

A typical $T$-$E$ phase diagram of squaric acid, exhibiting the two-step polarization reorientation, is presented in fig.~\ref{Smol} for the case of the field $E_1$, directed along the crystallographic axis $a$.  These diagrams have been constructed for different field orientations and extensively discussed in Refs.~\onlinecite{moina:21,moina:21:2,moina:22}.  There are three phases in the diagram:  AFE*, non-collinear antiferrielectric with nearly antiparallel and partially uncompensated sublattice polarizations; FE,  collinear field-induced ferroelectric \footnote{The AFE* phase is nothing but a tweaked by the external field initial fully compensated collinear AFE phase. At low fields and high temperatures (above $T_{{\rm N}0}$) the FE phase is the initial paraelectric phase, in which the external field induces small polarization; the sublattice polarization vectors are equal and parallel. With increasing field and lowering temperature, circumventing lines IV and III, it gradually transforms into the fully ordered ferroelectric phase. The term ``polar'' could be also appropriate for this phase.}, and NC90, noncollinear ferrielectric, where the angle between the sublattice polarizations mostly remains close to 90$^\circ$. The phases are separated by the lines of the first order phase transitions I, II, and III, and of the second order phase transitions IV, each terminating at some critical end point (bicritical end points BCE, tricritical point TCP, or critical end point CEP).  At low temperatures, when the field is increased, one of the sublattice polarizations is switched twice by approximately 90$^\circ$ each time, first at line II, to become perpendicular to the other sublattice polarization, and then at line III to become parallel to it. In the region between the critical points BCE$_1$ and BCE$_2$, a crossover from the AFE* to NC90 phase occurs, at which the sublattice polarization rotates continuously with increasing field.

The theoretical values of the AFE*-NC90 switching fields and the $P_1(E_1)$ polarization curves and the experimental data \cite{horiuchi:21} are in a reasonable agreement, which improves significantly when the experimental samples are of a better quality and of the increased dielectric strength \cite{moina:22}.  The NC90-FE transition has not been detected experimentally yet, as the maximal magnitudes of the electric field  that can be safely applied to the best quality crystals of squaric acid \cite{horiuchi:21} are still below 230~kV/cm. Also,  the calculated fields of the NC90-FE transition strongly depend on the choice of the model parameter values \cite{moina:21,moina:21:2,moina:22}. Therefore, the position of line III in the phase diagram (fig.~\ref{Smol}) is nominal and subject to change, should new experimental data become available, allowing us to ascertain the values of the fitting parameters.

The motivation to explore the ECE in squaric acid is thus
two-fold. First, because of the existence of the unique intermediate noncollinear ferrielectric phase NC90. It is interesting to determine the sign of the ECE in this phase and to study its field and temperature variations, as they are likely to be different from those in the AFE* and FE phases. Second, because of the presence of two bicritical end points in the field range below the limit of the experimental dielectric strength of the crystal. The ECE is expected to be strongly enhanced in the vicinity of the critical end points.

In the next Section we present some basic formulae, used
for description of the ECE. Then we proceed to explore the $T$-$E$ dependences of the entropy and polarization of squaric acid, and then to the ECE itself. The expressions for the entropy and polarization, obtained using the proton ordering model \cite{moina:21,moina:20}, are given in the Appendix.

\section{The calculations}
By definition, the adiabatic EC temperature shift is
the change of the system temperature from $T^{init}$ to
$T^{end}=T^{init}+\Delta T$ at changing the external applied field from 
$\textbf{E}^{init}$ to $\textbf{E}^{end}$, while the  (molar) entropy of the system
is kept constant
\be
S(T^{init}+\Delta T,\textbf{E}^{end})=
S(T^{init},\textbf{E}^{init}).
\label{deltat}
\ee
If $S={}$const, and external pressure is assumed to be zero, then from
\[
dS=\left(\frac{\partial S}{\partial \textbf{E}}\right)_T d\textbf{E}+\left(\frac{\partial S}{\partial T }\right)_{\textbf{E}}dT=0,
\]
it follows that
\be
dT=T\bm{\Gamma}_E d\textbf{E}.
\ee
We introduce here the electric Gr\"{u}neisen parameter
\be
\label{gru}
\bm{\Gamma}_{E}=-\frac 1T\left(\frac{\partial S}{\partial \textbf{E}}\right)_T\left(\frac{\partial S}{\partial T }\right)_{E}^{-1}=-\frac{V_m}{C_E}\left(\frac{\partial \textbf{P}}{\partial T }\right)_E
\ee
as a differential characteristic of the ECE, as opposed to $\Delta T$ being the integral one. Here $V_m$ is the molar volume; $C_{E}=T\left(\frac{\partial S}{\partial T }\right)_{E}$ is the molar heat capacity at constant electric field, and the Maxwell relation
\[
\left(\frac{\partial S}{\partial \textbf{E}}\right)_T=V_m\left(\frac{\partial \textbf{P}}{\partial T }\right)_E
\]
was used. Efficiency of the EC heating/cooling is sometimes characterized by the ECE strength or EC responsivity
\be
\frac{\Delta T}{\Delta E}=\frac{T^{end}-T^{init}}{|\textbf{E}^{end}-\textbf{E}^{init}|}.
\ee


We shall consider the case of the electric field $E_1$ applied along the $a$-axis.
We start from analyzing the $(T,E)$ landscapes of the molar entropy $S$ and polarization $P_1$ of squaric acid, calculated using the previously developed model \cite{moina:21}. A brief description of the model and the obtained expressions for the entropy, Eq. (\ref{Stot}), and polarization, Eq. (\ref{ppm}), are given in Appendix. All the relevant details, discussions,  formulae, fitting procedure, and values of the model parameters  can be found in Refs.~\onlinecite{moina:20,moina:21,moina:21:2,moina:22}.  The entropy is differentiated numerically with respect to temperature and field to find the component of the Gr\"uneisen parameter vector $\Gamma_{1E}$, hereafter referred to as simply the Gr\"uneisen parameter. To determine the EC temperature shift we solve Eq.~(\ref{deltat}) with respect to $\Delta T$ at different $T_{init}$ and $E_1^{end}$ and fixed $E_{1}^{init}=0$.

\begin{figure}[hbt]
\textit{}	\includegraphics[height=0.5\textwidth]{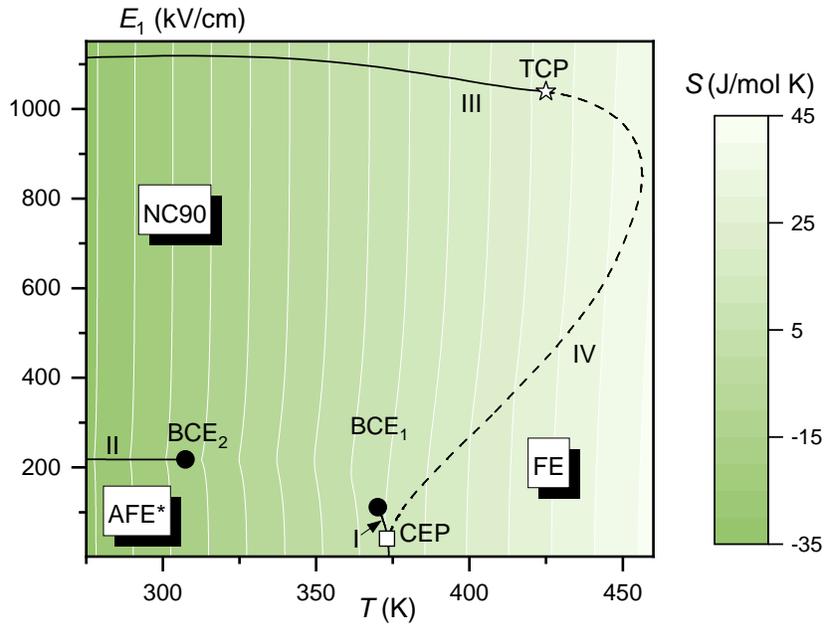}
	\caption{ The $T$-$E_1$ phase diagram of squaric acid taken from Ref.~\onlinecite{moina:21}, overlapping the contour color plot of the molar entropy $S$. Solid and dashed black lines indicate the first and second order phase transitions, respectively. The open square {\large $\square$}, star \FiveStarOpen, and full circles {\large $\bullet$} indicate the critical end point (CEP), tricritical point (TCP), and bicritical end points (BCE), respectively. The white lines are the contour lines of constant entropy. }
	\label{Smol}
\end{figure}

\begin{figure}[hbt]
	\includegraphics[width=0.7\textwidth]{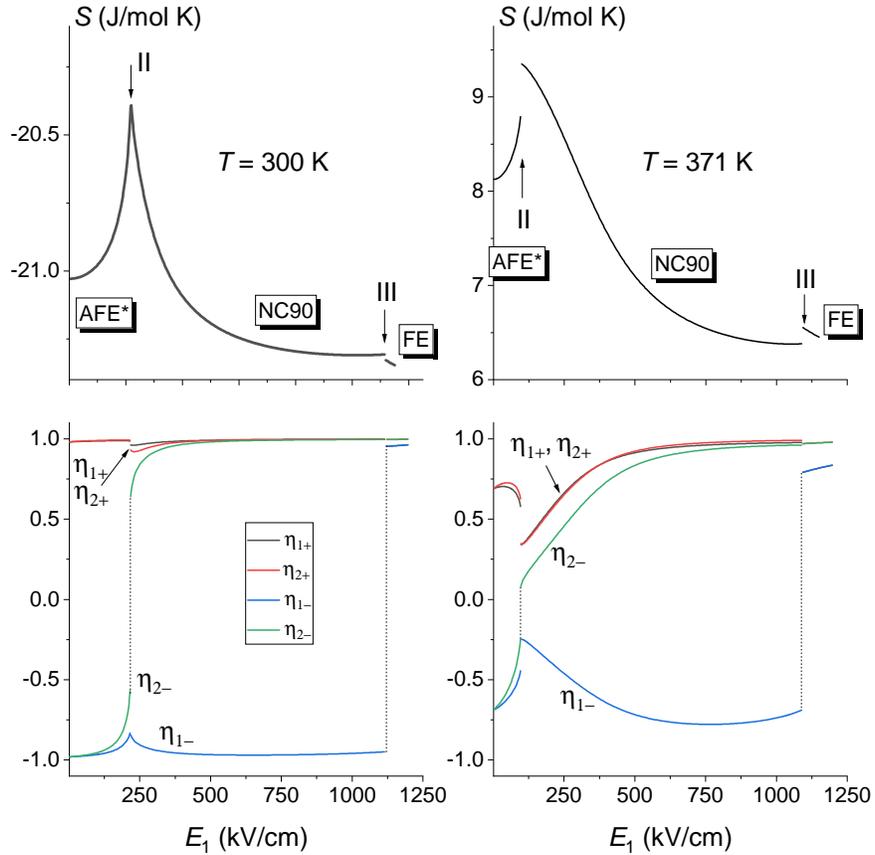}
	\caption{ The field dependences of the molar entropy $S$ (top)  and order parameters $\eta_{f\pm}$ (bottom) at 300 and 371~K. The first order transitions at lines I, II, and III from the $T$-$E_1$ phase diagram (fig.~\ref{Smol}) are indicated by arrows.  Vertical dotted lines indicate the jumps of the order parameters $\eta_{1-}$ and $\eta_{2-}$ at the first order transitions. The jump of entropy at line II is too small to be discernible.}
	\label{ordparentr}
\end{figure}

The contour plot of the molar entropy is presented in fig.~\ref{Smol}. Bending of the isentropic contour lines towards lower temperatures indicate the entropy accumulation at the
transition lines I and II and at the crossover between the AFE* and NC90 phases. The entropy increases with the field in the AFE* phase and decreases in FE and most of NC90 phase. According to Eq.~(\ref{gru}) this means negative and positive Gr\"uneisen parameter $\Gamma_{1E}$, respectively, or, in other words,  lowering temperature down by the increasing field $E_1$ in the AFE* phase and raising it up in the FE  and NC90 phases. A better insight can be obtained from the analysis of the field dependence of the order parameters, illustrated in fig.~\ref{ordparentr}.

As described in Appendix, the system behaviour is governed by four order parameters $\eta_{1\pm}$ and $\eta_{2\pm}$, two for each sublattice, `+` and `$-$`. The system is most ordered and has the lowest entropy, when the absolute values of the order parameters are close to 1. This can be confirmed by plotting the  entropy, Eq.~(\ref{Stot}), as a function of, say, $\eta_{1-}$ and $\eta_{2-}$ at constant temperature and electric field. As seen in the bottom panel of fig.~\ref{ordparentr}, at approaching the first order transition at lines I and II from below, the absolute values of all order parameters decrease, which results in a marked increase of the entropy (negative $\Gamma_{1E}$). At the transition $\eta_{2-}$ jumps to a positive value, indicating the  rotation of the `$-$` sublattice polarization. Above the transition, the perfect ferrielectric ordering is approached, as all four $|\eta_{f\pm}|$ increase. The entropy decreases, and $\Gamma_{1E}$ becomes positive. 
%
The entropy behavior in the vicinity of the NC90-FE phase transition at line III is qualitatively similar, which means that $\Gamma_{1E}$ again becomes negative at approaching this line from below and positive above it. However, at fields this high, the system is almost completely ordered in both phases, especially at low temperatures. The entropy, therefore, remains low and barely changes, and the magnitudes of $\Gamma_{1E}$ in this region are expected to be low too.

The field variation of the order parameters and entropy at 371~K is stronger than at 300~K.
There is, however, a more important distinction.
At 300~K this variation is nearly symmetric/antisymmetric 
at approaching the transition field at line II from below and from above. At 371~K, on the other hand, it is highly asymmetric below and above line I. Note, for instance, that $|\partial S/\partial E_1|$ and $|\partial \eta_{f-}/\partial E_1|$ nearly diverge at approaching line II from both sides and line I from below (and do diverge at the endpoints BCE$_2$ and BCE$_1$ of these lines), but remain moderate as line I is approached from above. One should expect different behavior of the Gr\"uneisen parameter in the vicinities of lines I and II and a strong asymmetry of its field dependence near line I.

The entropy and ECE behavior of squaric acid under external electric field $E_1$ can then be summarized as follows. In the AFE* phase the field spoils the perfect compensated AFE ordering and increases the entropy ($\Gamma_{1E}<0$). In the NC90 phase just above the switching, the field increases the noncollinear ferrielectric ordering and decreases the entropy ($\Gamma_{1E}>0$). Much higher fields begin to destroy the perfect noncollinear ordering in favor of the collinear FE one and increase the entropy again ($\Gamma_{1E}<0$). In the FE phase
the field only increases the ordering and decreases the entropy ($\Gamma_{1E}>0$).

\begin{figure}[hbt]
	\includegraphics[height=0.5\textwidth]{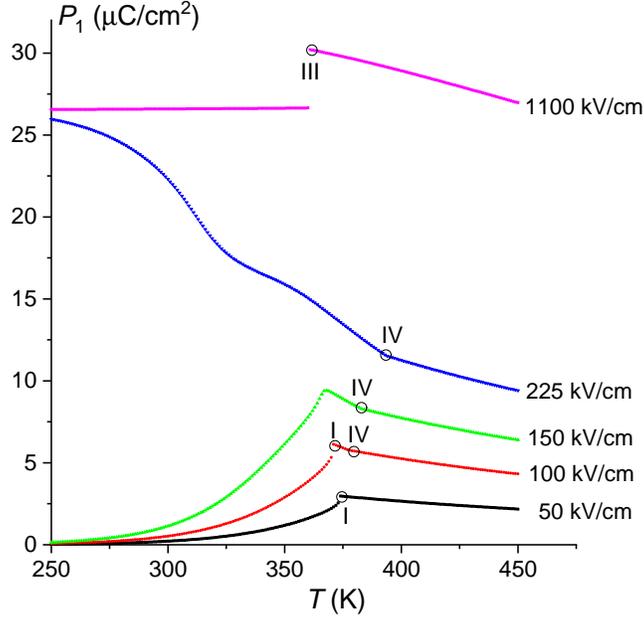}
	\caption{ 
		The temperature dependences of polarization $P_1$ at different values of the electric field $E_1$. Open circles show the phase transitions at lines I-IV. The rounded peak at 150~kV/cm corresponds to the crossover between the AFE* and NC90 phases.}
	\label{pol}
\end{figure}

The temperature curves of the polarization $P_1$, presented in fig.~\ref{pol}, lead to the same conclusions. The polarization increases with temperature in the AFE* phase (negative $\Gamma_{1E}$ expected) and decreases in FE and most of the NC90 phase (positive $\Gamma_{1E}$). At very high fields just below line III, the polarization is almost independent of the field, but begins again to slightly increase with temperature, implying $\Gamma_{1E}\lesssim 0$ here.

\begin{figure}[hbt]
	\includegraphics[height=0.5\textwidth]{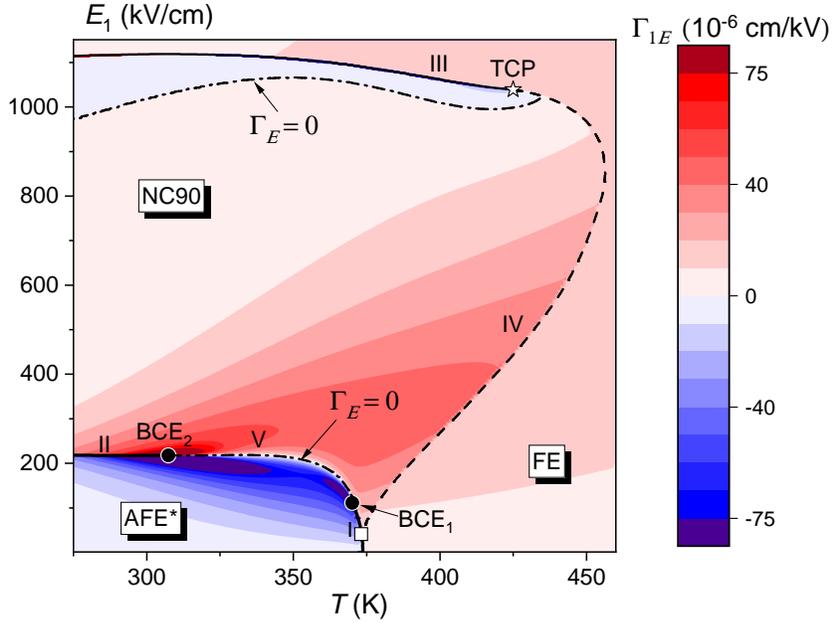}~~~
	\caption{\label{fig-gru} The $T$-$E_1$ contour plot of the Gr\"{u}neisen parameter.   The solid and dashed lines, as well as the symbols are the same as in fig.~\ref{Smol}. Dash-dotted lines indicate the change of sign of $\Gamma_E$.}
\end{figure}

The $T$-$E_1$ landscape of the Gr\"{u}neisen parameter  $\Gamma_{1E}$ in fig.~\ref{fig-gru} agrees with the qualitative predictions, following from the analysis of the entropy and polarization.
As one can see, the ECE  is indeed negative ($\Gamma_{1E}<0$) within the AFE* phase. At lines I and II of the first order phase transition to the FE and noncollinear NC90 phases, the Gr\"uneisen parameter becomes positive.  In the NC90 phase $\Gamma_{1E}$ mostly stays positive, but decreases with increasing field. Within the narrow stripe along the line III it becomes negative again.
In the FE phase $\Gamma_{1E}>0$ at any temperature or field.

In the crossover region between the AFE* and NC90 phases the Gr\"uneisen parameter changes its sign at line V. As a continuation of the transition lines I and II beyond the end points BCE$_1$ and BCE$_2$, this line represents yet another way to mark the crossover between the two phases. Frequently, the supercritical lines are drawn using the loci of the extrema of the response functions -- second derivatives of the thermodynamic potential (susceptibility, compressibility, \textit{etc}). In this case, the condition $\Gamma_{1E}=0$, according to Eq.~(\ref{gru}), means that $(\partial S/\partial E)_T=0$ and $(\partial P/\partial T)_E=0$. In other words, the maxima of the observables, i.e. the first derivatives of the thermodynamic potential: entropy isotherms or polarization isofields, are used.

The largest absolute values of $\Gamma_{1E}$ are predicted to occur in the vicinities of the bicritical end points BCE$_1$ and BCE$_2$  in the AFE* phase, and also  next to BCE$_2$ in the NC90 phase, where $\partial S/\partial E_1\to \infty$, but not  next to BCE$_1$ in the NC90 phase, where $\partial S/\partial E_1$ remains finite (see fig.~\ref{ordparentr}). This confirms the predicted earlier different behavior of the Gr\"uneisen parameter near the two endpoints BCE$_1$ and BCE$_2$ and its asymmetry  below and above the switching field near BCE$_1$.

\begin{figure}[hbt]
	\includegraphics[width=0.7\textwidth]{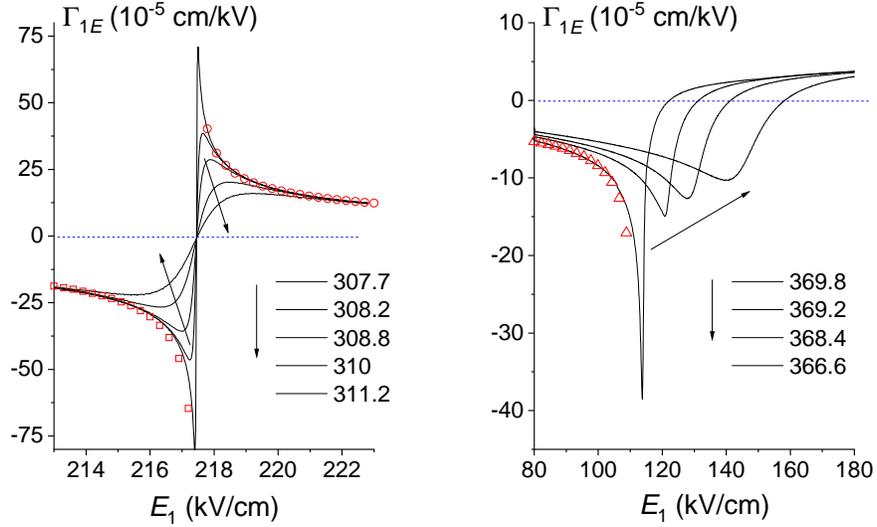}
	\caption{Dependences of the  Gr\"{u}neisen parameter on the field $E_1$ at different temperatures in the vicinities of the BCE$_2$ (left) and BCE$_1$ (right) points. Lines: the theory. Symbols:  the power law  fits according to Eq.~(\ref{scaling}) with  $G=3.6\cdot 10^{-5}$~cm/kV  ($\square$), $2.6\cdot 10^{-5}$~cm/kV  ($\circ$), and $3.1\cdot 10^{-5}$~cm/kV ($\bigtriangleup$).}
	\label{grun-scal}
\end{figure}

Both findings are further illustrated by the shown in fig.~\ref{grun-scal}  field dependences of the Gr\"uneisen parameter $\Gamma_{1E}$ in the crossover region. Near BCE$_2$ (the left panel) all curves, corresponding to different temperatures, cross at approximately the same ``isosbestic'' point, where $\Gamma_{1E}=0$. It reflects the fact that line V near BCE$_2$ is almost parallel to the temperature axis, i.e. the crossover field is barely temperature dependent. Near BCE$_1$ (the right panel), the crossover field strongly increases with decreasing temperature. The field dependences of the Gr\"uneisen parameter near BCE$_2$ are nearly antisymmetric. The sharp negative and positive peaks below and above the crossover point are smeared out with increasing temperature, as the system moves away from BCE$_2$. On the other hand, the field dependences of $\Gamma_{1E}$ near BCE$_1$ are highly asymmetric. Negative peaks are observed below the crossover field, similarly smeared out, as the system moves away from BCE$_1$. On the other hand,  above the crossover field $\Gamma_{1E}$ monotonically increases with $E_1$ into a rounded shoulder. 

The tails of all three peaks in fig.~\ref{grun-scal} at $T$ closest to $T_{\textrm{BCE}}$ are well approximated by the power law dependences 
\be
\label{scaling}
\Gamma_{1E}= \text{sgn}(e-1) G|e-1|^{-\alpha},
\ee
where $e$ is the applied field normalized per the coordinate
of the respective end point BCE$_1$ or BCE$_2$
\[
e=\frac{E_1}{E_{\textrm{BCE}}}.
\]
Those coordinates are quite sensitive to the choice of the model parameters values. At the used in the present calculation ones \cite{moina:21,moina:21:2} they are:
$T_{\textrm{BCE2}}=307.4$~K,	$E_{\textrm{BCE2}}=217.4$~kV/cm,
$T_{\textrm{BCE1}}=370.1$~K,	$E_{\textrm{BCE1}}=111$~kV/cm.
The exponent $\alpha$ was found to be about 0.42 both below and above BCE$_2$ and 0.48 below BCE$_1$. The coefficients $G$ are different for each branch and given in the caption to fig.~\ref{scaling}. The $G_{1E}(E_1)$ curves at larger $|T-T_{\textrm{BCE}}|$ approach the power law dependences (\ref{scaling}) asymptotically with increasing $|e-1|$.


It has been discussed earlier~\cite{moina:21} that the noncollinear ferrielectric phase within the narrow wedge between lines I and IV above CEP, may, in fact, be different from the NC90 phase above line II. The differences between the field behaviour of the order parameters, entropy, and the Gr\"uneisen parameter above line I and above line II seem to give further evidence for this assumption. There is also a possibility that the system of the critical end points BCE$_1$ and CEP is an artefact of the mean field approximation \cite{plascak:03} used for the long-range interactions, and in a more advanced approximation these end points should merge into a single tricritical point. In that case, the wedge between lines I and IV disappears, and the system behavior in this region will be  different. Additional experimental measurements and  Monte-Carlo calculations could shed some light on this problem.

	\begin{figure}[hbt]
	\includegraphics[height=0.5\textwidth]{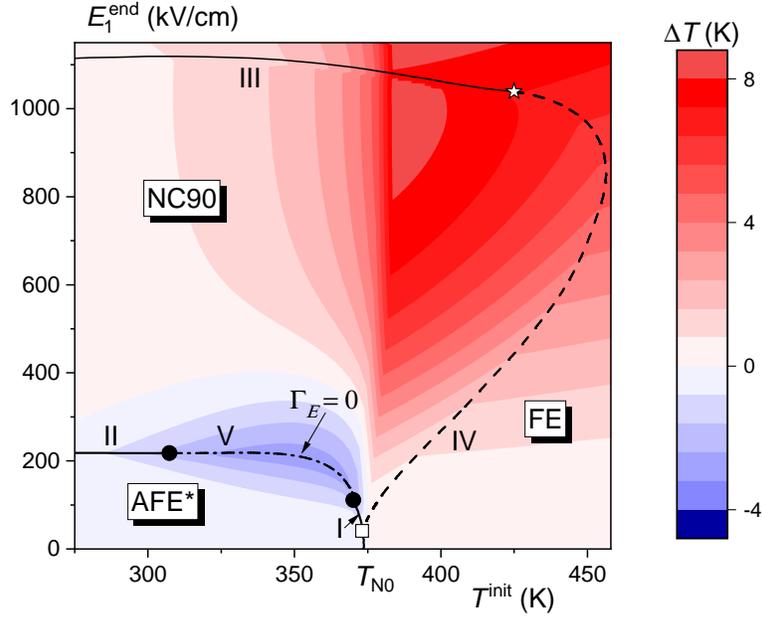}
	\caption{\label{fig-deltat} The $T^{\rm{init}}$-$E^{\rm {end}}$ contour plot of the adiabatic ECE temperature shift by the field $E_1$ (right). $E_1^{\rm{init}}=0$. The solid and dashed lines, as well as the symbols are the same as in fig.~\ref{Smol} and correspond to $T^{init}$. }
\end{figure}

Finally, the $T^{init}$-$E_1^{end}$ contour plot of the EC temperature shift $\Delta T$  is given in fig.~\ref{fig-deltat}. 
Since it is calculated for $E_{1}^{init}=0$, the entire landscape bears the imprint of the system entropy behavior at zero field, revealed as the jumps in $\Delta T$ at $T^{init}=T_{{\rm N}0}$ at all fields. 
Note that other anomalies of $\Delta T$, visualized, for instance, as bends of the contour lines, occur at $T^{end}$ and do not coincide with the phase transition lines, drawn for $T^{init}$.

In the AFE* phase the EC temperature shift $\Delta T$
is negative, and its magnitude increases with the end field $E_1^{end}$. At  $T^{init}>T_{{\rm N}0}$, it
is positive at all values of $E_1^{end}$, both in the FE and in NC90 phases.
Below  $T_{{\rm N}0}$ in the NC90 phase, the temperature shift first remains negative despite the change of sign of $\Gamma_E$ at lines I, II, and V. However, its amplitude decreases with increasing field $E_1^{end}$, until the effect of positive $\Gamma_E$ compensates the negative shift accumulated in the AFE* phase. At $E_1^{end}$ above 400 kV/cm $\Delta T$ is positive at all temperatures.

The largest negative EC temperature shifts $\Delta T$, up to $-2.7$~K, are predicted to be achieved in the vicinity of the crossover line V at around 200~kV/cm,  i.e. with the experimentally attainable fields \cite{horiuchi:21}. The respective maximal ECE strength $\Delta T/\Delta E$ is about 0.013~K cm/kV. Positive EC temperature shifts do not exceed 1~K at $E_{1}^{end}$ below 250~kV/cm.
It is predicted that $\Delta T$ can reach 8~K and more, however, the fields of the order of 800~kV/cm are required for that, which is well beyond the obtained so far largest dielectric strength of squaric acid crystals \cite{horiuchi:21}.

\begin{figure}[hbt]
	\includegraphics[width=0.5\textwidth]{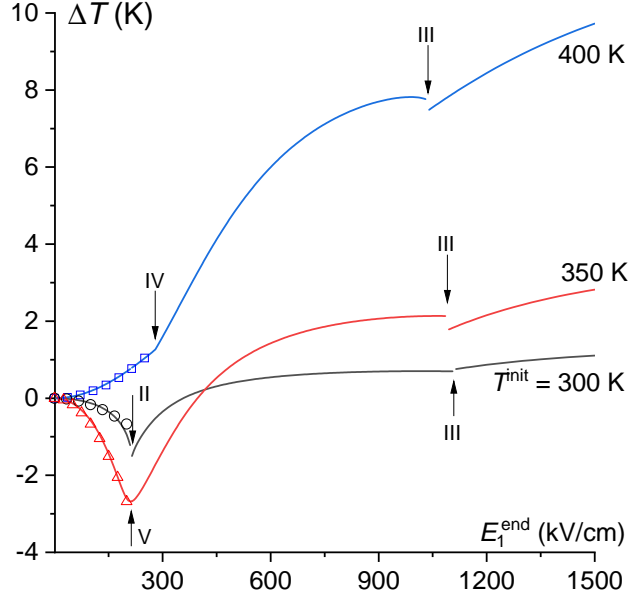}
	\caption{The field dependence of the adiabatic EC temperature shifts $\Delta T$ at  $E_1^{init}=0$ and different $T^{init}$. Lines: the theory. Symbols:  quadratic fits $\Delta T=k(E_1^{end})^2$, where $k=-1.7 \cdot 10^{-5}$~K cm$^2$/kV$^2$ for $T^{init}=300$~K ($\bigcirc$), $-6.7 \cdot 10^{-5}$ for 350~K ($\bigtriangleup$), and $1.7 \cdot 10^{-5}$ for 400~K ($\square$). Anomalies of $\Delta T$ in the vicinities of the phase transition lines II, III, IV and of the crossover line V are indicated by the arrows. }
	\label{ECE-scal}
\end{figure}

The field dependence of the adiabatic EC temperature shifts at different $T^{init}$ is shown in fig.~\ref{ECE-scal}. At low fields $\Delta T$ can be well approximated by quadratic functions $\Delta T =k (E_{1}^{end})^2$ for $T^{init}$ both below and above $T_{{\rm N}0}$. The temperature dependent coefficient $k$ is negative for $T^{init}<T_{{\rm N}0}$ and positive for $T^{init}>T_{{\rm N}0}$. This is in agreement with the predicted scaling law for the EC temperature shift in antiferroelectrics \cite{lisenkov:2016}.  Just above the transition (crossover) to the NC90 phase, $\Delta T$ is a concave down function of $E_{1}^{end}$ at 300~K, concave up at 350~K, and a linear function at 400~K.  Small jumps of $\Delta T$ are observed at crossing the first order transition lines II and III. The jump directions are opposite to the directions of the respective polarization $P_1$ jumps. In the high-field
region of the FE phase (above line III of the phase diagram) $\Delta T$ is a weakly nonlinear, concave down function of $E_{1}^{end}$.

\section{Concluding remarks}
In the present paper we study the electrocaloric effect in squaric acid, two-dimensional hydrogen-bonded pseudotetragonal antiferroelectric. The case of the electric field $E_1$ applied along the crystallographic $a$ axis of the crystal is considered. The $T$-$E_1$ phase diagram of this crystal is quite rich \cite{moina:21}, exhibiting
three phases, including the noncollinear ferrielectric one (NC90) with perpendicular sublattice polarizations, as well the first and second order transitions, critical end points, etc. The field-induced polarization switching is expected to be an unusual two-stage process, with the sublattice polarization vector rotated by $90^\circ$ at each stage, and the NC90 phase being the intermediate one between the AFE* and FE phases. 

The calculations are performed using the previously elaborated \cite{moina:20,moina:21,moina:21:2,moina:22} deformable proton-ordering model, which has proved to be successful for description of the squaric acid behavior at ambient conditions, under hydrostatic pressure, and in external electric fields, applied within the plane of the hydrogen bonds.
The $T$-$E_1$ landscapes of the system entropy and polarization are explored. The electric Gr\"uneisen parameter $\Gamma_{1E}$ and ECE temperature shift $\Delta T$, being the local in temperature/field and the integral characteristics of the ECE, respectively, are calculated.

It has been found that the ECE, as expected, is negative ($\Gamma_{1E}<0$) in the AFE* phase, where the field destroys the perfect AFE order and increases the entropy. The ECE is positive ($\Gamma_{1E}>0$) in the FE phase, where the field enhances the order and decreases the entropy. In the intermediate noncollinear phase NC90, the ECE is positive at fields just above the switching from the AFE* phase, where the field increases the noncollinear ferrielectric order. With increasing field  $\Gamma_{1E}$ gradually decreases and becomes negative again at  fields below the switching to the FE phase, where the field begins to destroy the perfect noncollinear ordering in favor of the collinear FE one. 

At low fields, the EC temperature shift is proportional to $(E_{1}^{end})^2$ at all temperatures, both in the AFE* and in the FE phases. The largest calculated EC temperature shift below 230~kV/cm, which is the best obtained to date dielectric strength of squaric acid samples \cite{horiuchi:21}, is expected to occur along this line, reaching $-2.7$~K at 200~kV/cm with the EC responsivity of $-0.013$~K cm/kV.

Most interesting are the peculiarities of $\Gamma_{1E}$ caused by the presence of two bicritical end points BCE$_1$ and BCE$_2$ in the phase diagram.
The line, connecting BCE$_1$ and BCE$_2$, where $\Gamma_{1E}=0$ (line V in fig.~\ref{gru}), is suggested to
be used to mark the crossover between the AFE and NC90 phases. 
Behavior of $\Gamma_{1E}$ is found to be qualitatively different in the vicinities of  BCE$_1$ and BCE$_2$: practically antisymmetric near BCE$_2$ and completely asymmetric near BCE$_1$. It indicates that the noncollinear ferrielectric phase near BCE$_1$ might be different from the NC90 one, if only the BCE$_1$ endpoint is not the artefact of the used mean field approximation. Further experimental studies and Monte-Carlo calculations are required to clarify this situation.

	\section*{Data Availability Statement}
				The data that support the findings of this study are available from the corresponding author upon reasonable request.
		
		
	\appendix
		\section{}
The used deformable proton ordering model \cite{moina:20,moina:21} considers the motion of protons in two alternating plane sublattices, to be denoted hereafter as '+' and '$-$', placed in the thermally expanding host lattice of heavy ions that forms the double-well potentials for each proton. This motion is described by Ising pseudospins, whose two eigenvalues $\sigma=\pm 1$ are assigned to two equilibrium positions of each proton. The arrangement of pseudospins in each sublattice corresponds to the structure presented in fig.~\ref{structure}.

It is assumed that electric fields $E_1$ and $E_3$ are applied to the crystal along its $a$ and $c$ axes. These fields break the equivalence of the hydrogen bonds linking the C$_4$O$_4$ groups along these axes, as well as the equivalence of the two sublattices. That leaves us with four independent order parameters \cite{moina:21}
$ \eta=\langle \sigma \rangle$,
two for each sublattice:
$\eta_{1\pm}$ and $\eta_{2\pm}$.
In  absence of the electric fields we have
$\eta_{1+}=\eta_{2+}=-\eta_{1-}=-\eta_{2-}$. Polarization rotation by 90$^\circ$ in a fully ordered system is signalled by the change of the sign of one of the order parameters in the `$-$` sublattice.

The system Hamiltonian \cite{moina:21,moina:21:2} includes intra- and interplane long-range interactions between the pseudospins, ensuring ferroelectric ordering within each separate plane and alternation of polarizations in the stacked planes, as well as the short-range interactions, ensuring  localization of protons in the lateral configurations (see fig.~\ref{structure}). The mean field and four-particle cluster approximations are used for the long-range and short-range interactions, respectively.

In the short-range Hamiltonian the usual Slater-Takagi type scheme \cite{matsushita:80,matsushita:82,moina:20,moina:21} of 16 degenerate levels of lateral/diagonal/single-ionized/double-ionized proton  configurations is exploited. The degeneracy of the energy levels of the lateral and  single-ionized  configurations, possessing dipole moments in the $ac$ plane,  is removed by the electric fields $E_1$ and $E_3$. 
Assignment of the dipole moments to these configurations is a crucial point of the model, and it has been described earlier \cite{moina:21,moina:21:2} in great detail. For instance, the vector of the dipole moment, assigned to the H$_2$C$_4$O$_4$ group with protons in the lateral configuration
shown in fig.~\ref{structure} is \cite{moina:21} $(2\mu^H+2\mu_\parallel^\pi,0,-2\mu_\perp^\pi)$. This is based on the results of the Berry phase calculations \cite{horiuchi:18}, which have shown that the ground-state sublattice polarization in this crystal is formed directly by displacements of protons along the hydrogen bonds and, mostly, by the electronic contributions of switchable $\pi$-bond dipoles, and have indicated the directions for those contributions. 

From the presented in Refs.~\onlinecite{moina:21,moina:21:2} thermodynamic potential of the model system, its molar entropy is obtained straightforwardly
	\begin{widetext}
		\begin{eqnarray}
		&& \frac{\Delta S}{R}=\frac{\mathrm{v}}{2R}\sum_{ij=1}^3c_{ij}^{(0)}\alpha_i^{(0)}u_j +\frac12\left(\ln D_+ +\ln D_-\right) +\frac1{4}\left[\ln(1-\eta^2_{1+}) +\ln(1-\eta^2_{2+}) +
		\ln(1-\eta^2_{1-})+
		\ln(1-\eta^2_{2-})\right]\nonumber \\
		&&\label{sqa:tpot} {}- (\beta\nu-\frac{2\delta_T\lambda_\delta\nu_0}{k_{\rm B}})\left[\frac{(\eta_{1+}-\eta_{1-})^2}4+
		\frac{(\eta_{2+}-\eta_{2-})^2}4\right]-
		(\beta\nu'-\frac{2\delta_T\lambda_\delta\nu'_0}{k_{\rm B}})\left[\frac{(\eta_{1+}+\eta_{1-})^2}4+
		\frac{(\eta_{2+}+\eta_{2-})^2}4\right] \nonumber\\
		&&
		+\left(\beta-\frac{2\delta_T}{k_{\rm B}\lambda_\delta}\right)\left[
		\frac{a\eps+2bw(\cosh{z_{1+}}+\cosh{z_{2+}})}{2D_+}+
		\frac{a\eps+2bw(\cosh{z_{1-}}+\cosh{z_{2-}})}{2D_-}
		\right]\\&&   
		-\beta \sum_{f=1}^2\frac{\eta_{f+}+\eta_{f-}}4
		(\mu_{f1}E_1+\mu_{f3}E_3), \nonumber
		\end{eqnarray}
		where $R$ is the gas constant, $\beta=(k_\textrm{B}T)^{-1}$, and
		\begin{eqnarray}
		& D_{\pm}=a+\cosh(z_{1\pm}+z_{2\pm})+2b(\cosh z_{1\pm}+\cosh z_{2\pm})+\cosh(z_{1\pm}-z_{2\pm}),\nonumber \\
		\label{z}		& z_{f\pm}=\frac12\ln\frac{1+\eta_{f\pm}}{1-\eta_{f\pm}}\pm\beta\nu\frac{\eta_{f+}-\eta_{f-}}{2} + \beta\nu'\frac{\eta_{f+}+\eta_{f-}}{2} 
		+
		\beta\frac{\mu_{f1}E_1}{2}+\beta\frac{\mu_{f3}E_3}{2} , \quad (f=1,2); \\
		& a=\exp(-\beta\eps), \quad b=\exp(-\beta w),\nonumber\\
		& \mu_{11}=\mu_{23}=\mu^H+\mu^\pi_\parallel-\mu^\pi_\perp, \quad
		-\mu_{13}=\mu_{21}=\mu^H+\mu^\pi_\parallel+\mu^\pi_\perp .\nonumber
		\end{eqnarray}\end{widetext}
	The first term in (\ref{sqa:tpot}) comes from the ``seed'' energy of thermal expansion \cite{moina:20} of the host lattice;  $u_1$, $u_2$, and $u_3$ are the uniform diagonal lattice strains; $c_{ij}^{(0)}$ and $\alpha_i^{(0)}$ are the ``seed'' elastic constants and thermal expansion coefficients; $\mathrm{v}$ is the unit cell volume just above the AFE transition temperature $T_{{\rm N}0}$ in absence of the electric fields.

	The interaction parameters of the deformable model are temperature dependent. Thus, the short-range Slater-Takagi type energy parameters $	\eps$ and $w$  are considered \cite{moina:20} to be quadratic functions of the H-site distance $\delta$.  In its turn, the distance $\delta$ is taken to vary according to its experimentally observed \cite{semmingsen:95} above the transition  linear temperature dependence
	\begin{equation}
	\label{delta-model}
	\delta=\delta_0[1+\delta_T(T-T_{\textrm{N0}})]\equiv \delta_0\lambda_\delta,
	\end{equation}
	where $T_{\textrm{N0}}$ is the transition temperature at zero electric field. It yields
	\[
		\label{kdp-Slater}
	 \eps=\eps'_0\lambda_\delta^2, \quad w=w_0\lambda_\delta^2.
\]
	
	The long-range interaction parameters $\nu$ and $\nu'$ from Eq.~(\ref{sqa:tpot}) are taken to  depend on $\delta$ and on the lattice strains as \cite{moina:20,moina:21}
	\[
		 \nu=\nu_0\lambda_\delta^2+\sum_{i=1}^3\psi_iu_i,
		\quad
		\label{kdp-long}  \nu'=\nu'_0\lambda_\delta^2+\sum_{i=1}^3\psi_i'u_i.
\]
		Finally, the dipole moments $\mu_{f1}$ and $\mu_{f3}$ 
	are assumed to be temperature or strain independent. 
	
\begin{figure}[hbt]
	\includegraphics[height=0.5\textwidth]{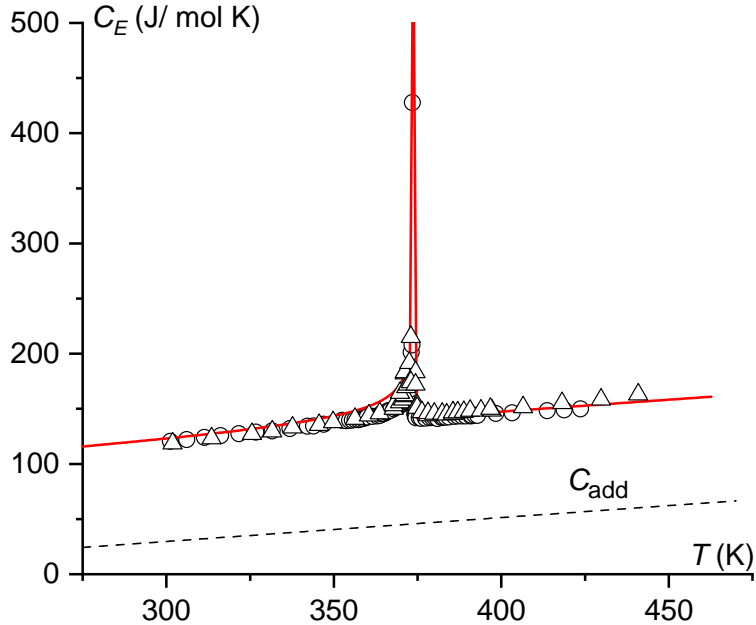}
	\caption{ The temperature dependence of the specific heat of squaric acid at zero electric field. Symbols: experimental points, taken from Refs.~\onlinecite{barth:79} ($\bigcirc$) and \onlinecite{gronvold:76} ($\triangle$). Dashed line: the added regular part of the specific heat, $C_{add}$.}
	\label{cmol}
\end{figure}

 Differentiating Eq.~(\ref{sqa:tpot}) numerically, we obtain the  contribution to the molar  specific heat of the crystal  from its subsystem, described by the model Hamiltonian
\[
\Delta C_E=T\left(\frac{\partial \Delta S}{\partial T}\right)_E.
\]
To get the total molar specific heat of the crystal that can be compared to experimental data, the background regular contribution from the host lattice has to be added to $\Delta C_E$ 
\[
C_E=\Delta C_E + C_{add}.\]
A sufficiently good agreement with the experiment, as illustrated in fig.~\ref{cmol}, is obtained if the added regular part 
is a linear function of temperature
	\[
	C_{add}=k_0+k_1 T,\]
with $k_0 = 45.62$~J/(mol K), $k_1=0.217$~J/(mol K$^2$).
One can also see, that $\Delta C_E$ does not represent a purely anomalous part of the total specific heat, but has a significant regular part as well, which, actually, is larger than $C_{add}$.

Integrating $C_{add}/T$ over temperature we find the added contribution to the entropy
\[
S_{add}=k_1(T-T_{{\rm N}0})+ k_0\ln (T/T_{{\rm N}0}) + \textrm{const} .\]
The total entropy, to be used to explore the ECE in squaric acid, is  
\be
\label{Stot}
S=\Delta S + S_{add}.
\ee 
Note that $S_{add}$ and $S$ are determined within a constant, which in the calculations is chosen to yield $S_{add}(T_{{\rm N}0})=0$.

The net polarization $\mathbf{P} $ of squaric acid has been obtained earlier
\cite{moina:21,moina:22}. It is a sum of the  sublattice polarization vectors $\mathbf{P}_\pm$
\begin{equation}
\label{ppm}
\mathbf{P}=\mathbf{P}_++\mathbf{P}_-,\quad \mathbf{P}_\pm=(P_{1\pm},0,{P_{3\pm}}), 
\end{equation}
where \cite{moina:21}
\begin{eqnarray*}
&& P_{1\pm} = 
\frac{\mu^H+\mu^\pi_\parallel}{2{\textrm v}}
(\eta_{1\pm}+\eta_{2\pm}) + \frac{\mu_\perp^\pi}{2{\textrm v}}(\eta_{2\pm}-\eta_{1\pm}),\nonumber \\
&& P_{3\pm}=
\frac{\mu^H+\mu^\pi_\parallel}{2{\textrm v}}
(\eta_{2\pm}-\eta_{1\pm}) - \frac{\mu_\perp^\pi}{2{\textrm v}}(\eta_{2\pm}+\eta_{1\pm}).
\label{net}
\end{eqnarray*}
Naturally, in absence of the electric fields
$\mathbf{P}=0$.

The order parameters $\eta_{f\pm}$ and lattice strains $u_i$ are found by numerical minimization of the presented in Refs.~\onlinecite{moina:21,moina:21:2} thermodynamic potential of the system, using the given therein values of the model parameters.
The fitting procedure has been  discussed earlier \cite{moina:20,moina:21,moina:22} very extensively.



\begin{thebibliography}{10}
	
	 \bibitem{scott:16}
	Y. Liu, J.F. Scott, and  B. Dkhil,
	Appl. Phys. Rev. \textbf{3}, 031102  (2016) \doi{10.1063/1.4958327}.
	
	 \bibitem{moya:14}
	X. Moya, S. Kar-Narayan, and  N. D. Mathur,  Nature Materials  \textbf{13}, 439 (2014) \doi{10.1038/nmat3951}.
	
		\bibitem{kutnjak:15}
	Z. Kutnjak, B. Ro\v{z}i\v{c}, and R. Pirc,``Electrocaloric Effect: Theory, Measurements,
	and Applications'', in \textit{Wiley Encyclopedia of Electrical
		and Electronics Engineering}, edited by J. G. Webster (Wiley, Hoboken, 2015), p. 1. \doi{10.1002/047134608X.W8244}.
	
	
	\bibitem{grunebohm:18}
	A. Gr\"{u}nebohm, Y.-B. Ma, M. Marathe, B.-X. Xu \textit{et al},
	Energy Technol. \textbf{6}, 1491 (2018)
	\doi{10.1002/ente.v6.8}.
	

\bibitem{pirc:14}
R. Pirc, B. Ro\v{z}i\v{c}, J. Koruza, B. Mali\v{c}, and  Z. Kutnjak,
EPL, \textbf{107},  17002 (2014) \doi{10.1209/0295-5075/107/17002}
	
\bibitem{squillante:21}
L. Squillante, I.F. Mello, A.C. Seridonio, and M. de Souza, Materials Research Bulletin 
\textbf{142},  111413 (2021) \doi{10.1016/j.materresbull.2021.111413}	
	

	
	\bibitem{semmingsen:77}
	D. Semmingsen,  F. J. Hollander, and T. F. Koetzle, 
	J. Chem. Phys. \textbf{66}, 4405 (1977). \doi{10.1063/1.433745}
	
	\bibitem{semmingsen:95}
D. Semmingsen, Z. Tun,  R.J. Nelmes,  R. K. McMullan, and T. F. Koetzle,  Zeit. f\"{u}r Kristallographie \textbf{210}, 934 (1995) \doi{10.1524/zkri.1995.210.12.934}.
		
	
	\bibitem{horiuchi:18}
	S. Horiuchi, R. Kumai, and S. Ishibashi,  Chem. Sci. \textbf{9}, 425 (2018) \doi{10.1039/C7SC03859C}
	
	
	\bibitem{ishibashi:18}
	S. Ishibashi, S. Horiuchi, and R. Kumai,  Phys. Rev. B \textbf{97}, 184102 (2018) \doi{10.1103/PhysRevB.97.184102}
	
	
		\bibitem{horiuchi:21}
	S. Horiuchi,  S. Ishibashi, Chem. Phys.  \textbf{12}, 14198 (2021) \doi{10.1039/d1sc02729h}
	
	
		\bibitem{moina:21}
	A. P. Moina,  Phys. Rev. B  \textbf{103}, 214104 (2021) \doi{10.1103/PhysRevB.103.214104}
	
	
	\bibitem{moina:21:2}
	A. P. Moina, Condens. Matter Phys.  \textbf{24}, 
	43703 (2021) \doi{10.5488/CMP.24.43703}
	
	
		\bibitem{moina:22}
	A. P. Moina, Condens. Matter Phys. \textbf{25}, to be published (2022).
	
	\bibitem{moina:20}
	A. P. Moina,  Condens. Matter Phys. \textbf{23}, 33704 (2020) \doi{10.5488/CMP.23.33704}
	




\bibitem{lisenkov:2016}
S.~Lisenkov,  B.K. Mani, E. Glazkova, C. W. Miller, and I. Ponomareva, Sci. Rep. \textbf{6}, 19590 (2016)
\doi{10.1038/srep19590}.


\bibitem{plascak:03}
J. A. Plascak,  D. P. Landau, Phys. Rev. E   \textbf{67}, 015103(R) (2003) \doi{10.1103/PhysRevE.67.015103}.





\bibitem{matsushita:80}
E. Matsushita and  T. Matsubara,  Progr. Theor. Phys. \textbf{64}, 1176 (1980).


\bibitem{matsushita:82}
E. Matsushita and T. Matsubara,  Progr. Theor. Phys. \textbf{68},  1811 (1982).


\bibitem{barth:79}
 E. Barth,   J. Helwig,   H.-D. Maier,   H.E. M\"{u}ser, and  J. Petersson,
Z. Physik B  \textbf{34}, 393 (1979) 
\doi{10.1007/BF01325204}.

\bibitem{gronvold:76}
F. Gronvold, Pure and Appl. Chem. \textbf{47},  251 (1976). 




\end{thebibliography}

\end{document}